\documentclass[aps,prl,twocolumn,superscriptaddress]{revtex4-2}

\usepackage{xcolor}
\usepackage{multirow}
\usepackage{array}
\usepackage{makecell}
\usepackage{braket}
\usepackage{graphicx}
\usepackage{amsmath}
\usepackage{amssymb}
\usepackage{hyperref}
\usepackage{enumitem}
\usepackage{float}

\setcellgapes{3pt}
\makegapedcells

\begin{document}

\title{Reply to the Comment on ``The Axiom of Choice and the No-Signalling Principle''}

\author{Ämin Baumeler}
\affiliation{Faculty of Informatics, Università della Svizzera italiana,
Via la santa 1, Lugano-Viganello, 6962, Switzerland}

\author{Borivoje Daki\'c}
\affiliation{University of Vienna, Faculty of Physics, Vienna Center for Quantum Science and Technology, Boltzmanngasse 5, Vienna 1090, Austria}
\affiliation{Institute for Quantum Optics and Quantum Information (IQOQI), Austrian Academy of Sciences, Boltzmanngasse 3, Vienna 1090, Austria}

\author{Flavio Del Santo}
\affiliation{Constructor University, Bremen, Germany}
\affiliation{University of Geneva, Group of Applied Physics, Rue de lÉcole-de-Médecine 21, Geneva 1211, Switzerland}
\author{Milo\v s Milovanovi\' c}
\affiliation{Mathematical Institute of the Serbian Academy of Sciences and Arts, 
Kneza Mihaila 36, 11000 Belgrade, Serbia}
\begin{abstract}
We clarify the meaning of a probabilistic no-signalling strategy
in Ref.~\cite{baumeler2025axiom}. For every fixed input, the
deterministic strategy constructed using the Axiom of Choice can
indeed be represented by a Dirac probability distribution over
the outputs. The relevant distinction arises when one considers
the complete process from input to output, with the input itself
sampled according to a probability distribution. In this case,
the input and output spaces must be measurable and the dependence
of the conditional output distribution on the input must also be
measurable. Equivalently, the conditional distributions must form
a Markov kernel. The Axiom-of-Choice strategy fails precisely this
measurability requirement.
\end{abstract}

\maketitle

The main claim of the Comment~\cite{karvonen2026comment} is that
the deterministic strategy constructed in
Ref.~\cite{baumeler2025axiom} using the Axiom of Choice (AC) is in
fact a probabilistic no-signalling strategy, and that the
purported distinction between deterministic and probabilistic
no-signalling strategies therefore cannot hold. We clarify here
that, while the argument in the Comment is formally consistent
within its definitions, it does not capture the intended meaning
of ``probabilistic strategy'' in
Ref.~\cite{baumeler2025axiom}. We take this opportunity to make
explicit the definitions used there, which were not always stated
with sufficient precision.

The Comment defines, for each fixed input $x$, the empirical model
\begin{equation}
e_x=\delta_{g(x)},
\end{equation}
where $\delta_{g(x)}$ is the Dirac measure centered on the
deterministic output $g(x)$. Since each $e_x$ is individually a
probability measure over the outputs, the Comment concludes that
the family $(e_x)$ constitutes a probabilistic empirical model
according to Definition~2.2 of
Ref.~\cite{karvonen2026comment}.

This conclusion is correct under that definition. Once $x$ is
fixed, the strategy returns a definite output
$b\in\{0,1\}$, which can always be represented by a Dirac
probability distribution. Definition~2.2 of the Comment requires
only that a probability measure over the outputs is assigned to
every fixed input.

The notion intended in Ref.~\cite{baumeler2025axiom} is
operationally stronger. There, the input is itself sampled
according to a probability distribution. A probabilistic
strategy is therefore meant to describe a single stochastic
experiment in which a random input determines a conditional
probability distribution over the outputs.

Let $(X,\mathcal A)$ be the measurable input space and
$(Y,\mathcal B)$ the measurable output space. In the setting of
Ref.~\cite{baumeler2025axiom}, the input received by each party is
a real number in the unit interval, so that
\begin{equation}
X=[0,1],
\end{equation}
with $\mathcal A$ the Borel $\sigma$-algebra on $X$. The output
of each party is a bit,
\begin{equation}
b\in Y=\{0,1\},
\end{equation}
with $\mathcal B$ the discrete $\sigma$-algebra on $Y$.

For every fixed input $x\in X$, $e_x$ is a well-defined
probability distribution over the two possible outputs. To
describe the complete stochastic process, however, the
dependence of this distribution on the input must also be
measurable. In the terminology used in the Comment, the family
$(e_x)$ must form a Markov kernel \cite{klenke2008probability}.

This measurability condition is what gives the strategy its
operational meaning. In the game considered in
Ref.~\cite{baumeler2025axiom}, the input is sampled by the referee
according to a probability measure $\mu$. The conditional output
distributions must therefore be such that they can be averaged
over the random input. If $(e_x)$ is a Markov kernel, then for
every measurable output event $B\in\mathcal B$ one can define
\begin{equation}
\Pr(B)=\int_X e_x(B)\,d\mu(x).
\end{equation}
Thus, a Markov kernel describes a strategy that can be composed
with the referee's random choice of input to produce operational
probabilities for the game. A nonmeasurable pointwise family $(e_x)$
does not permit this composition.

For the deterministic strategy considered in
Ref.~\cite{baumeler2025axiom}, the referee samples an input
$x\in X$ according to a probability distribution, and the
strategy is described by a function
\begin{equation}
g:X\longrightarrow Y,
\end{equation}
which assigns the output $g(x)$ to the sampled input $x$.

In order to define a random variable, $g(x)$ must be measurable. Since the output space is binary,
this requires
\begin{equation}
g^{-1}(\{0\})\in\mathcal A
\qquad\text{and}\qquad
g^{-1}(\{1\})\in\mathcal A.
\end{equation}
Equivalently, for every $B\in\mathcal B$, the map
\begin{equation}
x\longmapsto e_x(B)
=
\delta_{g(x)}(B)
\end{equation}
must be measurable.

The AC-based strategy fails exactly this requirement. As discussed in
Ref.~\cite{baumeler2025axiom}, its complete input-output map is
not measurable. Thus, although it assigns a Dirac probability
distribution to every fixed input, these distributions do not
combine with the randomly sampled input to define a measurable
input-output process.

The Comment does not explicitly impose this measurability
condition in its definition of an empirical model. It observes
that one may ask why the distributions $e_x$ are not required to
depend measurably on $x$, thereby forming a Markov kernel, and
notes that this condition is trivial when the input space is
finite and discrete but becomes nontrivial in the infinitary
setting.

This is precisely the relevant distinction: In
Ref.~\cite{baumeler2025axiom},the term
``probabilistic strategy'' was intended to refer to the complete
stochastic process obtained when a randomly sampled input is
processed by the strategy.

The point of Ref.~\cite{baumeler2025axiom} is that no-signalling
can also be formulated without introducing probability
distributions or random variables, using only deterministic
functions. The AC-based strategy is well defined in this
functional framework. It also defines a probabilistic empirical
model in the pointwise sense adopted in the Comment, but not a
probabilistic strategy in the operational sense intended in
Ref.~\cite{baumeler2025axiom}.

The distinction intended in Ref.~\cite{baumeler2025axiom} is
therefore between functional no-signalling strategies and
probabilistic no-signalling strategies understood as measurable
stochastic input-output processes.

\bibliographystyle{unsrt}
\bibliography{references}

\end{document}